# Inverse thermal anisotropy in CdMgO measured using photothermal infrared radiometry and thermoreflectance

Misha Khalid[1], Ankur Chatterjee[1,2], Ewa Przezdziecka[3], Abinash Adhikari[3], Monika Stanke[1], Aleksandra Wierzbicka[3], Carlos J. Tavares[4], Michał Pawlak[1*]

[1]Institute of Physics, Faculty of Physics, Astronomy, and Informatics, Nicolaus Copernicus University, Grudziadzka 5, 87-100, Torun, Poland

[2]Chair of Applied Solid-State Physics, Experimental Physics VI, Ruhr-University Bochum, Universitaetsstrasse 150, D-44780, Bochum, Germany

[3]Institute of Physics, Polish Academy of Sciences, Al. Lotnikow 32/46, 02-668, Warsaw, Poland

[4]Centre of Physics of Minho and Porto Universities (CF-UM-UP), University of Minho, 4804-533 Guimaraes, Portugal

## Abstract

This study elucidates the intriguing phenomenon of inverse thermal anisotropy in cadmium magnesium oxide (CdMgO) thin films, characterized by cross-plane thermal conductivity being greater than in-plane thermal conductivity, essential for optimizing thermal management in next-generation optoelectronic devices. Herein, we utilized Photothermal Radiometry and Frequency Domain Thermoreflectance to precisely determine the thermal conductivity and diffusivity across various concentrations of magnesium in CdMgO alloys, thereby providing essential insights into thermophysical behavior. Atomic force microscopy and X-ray diffraction revealed a direct correlation between increasing magnesium content and progressive structural evolution within plasma-assisted molecular beam epitaxy-derived CdMgO alloys. Furthermore, heat transport mechanism, analyzed using Callaway and Abeles models, indicated key phonon interactions. This comprehensive investigation provides a framework for the precise control of CdMgO thin film thermal properties, paving the way for scalable fabrication strategies to optimize performance in high-power thermal management applications.

## 1. Introduction

The significant heat production of nanoscale systems underscores the necessity of regulating thermal characteristics, particularly in light of the rapid advancement of energy-efficient technology[1]. Efficient temperature management materials are essential in sectors such as thermoelectrics[2], optoelectronics[3], and microelectronics[4]. The directional dependence of thermal conductivity characterizes thermal anisotropy[5], a crucial element of this research that is crucial for improving energy conversion efficiency[6] and maximizing heat dissipation. Anisotropic thermophysical characteristics, such as thermal diffusivity ($\sigma$)[7], thermal conductivity ($k$), thermal effusivity ($e$)[8], and volumetric heat capacity ($Cv$)[9], were thoroughly investigated for materials including graphite[10], boron nitride (BN)[11], and silicon carbide (SiC)[12]. These characteristics determine the effectiveness of heat absorption, storage, and conduction in materials, making them crucial for applications needing special in-plane and cross-plane heat transfer[13].

A variety of strategies have been developed to evaluate thermophysical characteristics. These techniques include thermoreflectance in the frequency domains[14], spatial or time domain modulations (such as FDTR, SDTR, TDTR)[15], omega methods[16], scanning thermal microscopy (SThM)[17], photothermal radiometry (PTR)[18], the laser flash method[5], Raman spectroscopy[13], and the hot plate method[19]. Each method offers distinctive benefits suitable for particular material properties and measurement objectives. In anisotropic materials, the FDTR method is distinguished by its exceptional accuracy in frequency-domain measurements, which it employs to accurately measure directional heat flow[20]. The PTR approach is particularly advantage in that it measures thermal characteristics without the necessity of a metallic transducer layer, thereby reducing the potential for layer interactions to introduce uncertainty[21].

In previous work, Georges et al.[14] developed a PTR setup that integrates both frequency and spatial scans to achieve a resolution of 33 μm for measuring thermophysical properties such as anisotropic diffusivity without a transducer. They reported an anisotropic factor of 24.6 for flexible graphite and provided precise diffusivity measurements for a 1 μm titanium membrane. Chatterjee et al.[6] utilized FDTR with amplitude data to accurately measure cross-plane thermal conductivity, diffusivity, and thermal boundary resistance in a GaAs substrate. Their incorporation of deep learning techniques significantly enhanced analytical accuracy by addressing challenges related to local minima and optimizing fitting processes. Battaglia et al. [22] formulated one-dimensional and three-dimensional models for in-plane and cross-plane diffusivities in membranes, whereas Wang et al. [23] illustrated that the size of the pump beam influences measurement sensitivity. Huang et al., [24] demonstrate a scalable, maskless nanofabrication strategy to engineer surface nanocones on silicon, enhancing thermoelectric efficiency while preserving bulk crystallinity. Experimental measurements reveal a 40% reduction in thermal conductivity at room temperature, attributed to phonon surface scattering via an adjusted Callaway Holland framework across 4–295 K. Walwil et al.[25] present a transfer-based TDTR technique to measure thermal conductivity of porous thin films by depositing flat metal transducers on holey $SiO_2$, bypassing challenges of direct deposition. Experiments achieve <12% uncertainty, validated against effective medium theory, enabling reliable κ characterization for 13–50% porosity films with 1–3.5 μm pores. Epstein et al.[26] utilize frequency-domain thermoreflectance to probe pentacene film thermal conductivity and interfacial resistances on functionalized silicon, linking AFM-derived effective thickness to grain-size-dependent κ (77–300 K). Kapitza lengths (~150 nm) exceed film thicknesses, highlighting interface-driven thermal resistance critical for molecular electronics thermal design.

Recently, time-domain PTR utilizing periodic pulses was evaluated for amorphous thin films, demonstrating potential for nanoscale thermal diffusion research with accelerated infrared (IR) detectors. Similar to other oxides, CdMgO displays fascinating thermal characteristics, including an unusual inverse thermal anisotropy where cross-plane thermal conductivity surpasses in-plane conductivity[27]. This rare phenomenon presents challenges for thermal characterization and arises from factors such as crystal orientation or defect distributions. Gaining insights into this property is essential for optimizing devices based on CdMgO since anisotropy influences heat dissipation and overall device performance[7]. Research on materials like graphite and BN has shown that advanced methods such as FDTR and PTR especially when combining amplitude and phase data can effectively differentiate between in-plane and cross-plane thermal conductivities[28,29].

This study employs PTR and FDTR to experimentally investigate the thermal properties of CdMgO alloy films, with a particular emphasis on their inverse thermal anisotropy. Additionally, the Callaway[30] and Abeles models[31] are utilized to deepen the understanding of CdMgO heat transport behavior by providing a theoretical framework for analyzing lattice thermal conductivity. These models account for complex interactions among phonon scattering mechanisms, offering a robust approach to studying heat transport dynamics within the material. By integrating experimental findings with theoretical insights, this research aims to provide a comprehensive analysis of the heat transport phenomena in CdMgO, guiding efforts to optimize this material for advanced optoelectronic applications. This approach facilitates the development of tailored thermal management strategies that enhance the efficiency of technologies based on CdMgO. Key thermophysical properties measured include thermal diffusivity, effusivity, conductivity, and volumetric heat capacity findings that are crucial for advancing CdMgO application in high-performance devices requiring precise thermal control.

## 2. Experimental Methods

2.1 Material growth

CdMgO random alloys are generated using the plasma-assisted molecular beam epitaxy (PA-MBE) process on commercial quartz substrates. Dual-zone Knudsen effusion cells are used in the MBE system to supply magnesium (Mg) and cadmium (Cd), and an RF oxygen plasma cell is used to produce oxygen plasma. Acetone, methanol, and isopropanol are used in a 20-minute chemical cleaning procedure to thoroughly clean the quartz substrates before deposition. A thorough rinse with deionized water and a subsequent nitrogen gas drying procedure come next. The substrates were annealed for 1 hour in the load chamber at 100°C before the growing process. During the deposition phase, the fluxes of Cd and Mg are carefully regulated by fine-tuning the temperatures of the effusion cells. The Cd effusion cell temperature is maintained at 380°C throughout the procedure, which is carried out with a constant Cd beam equivalent pressure set at $2 \times 10^{-7}$. To create alloys with varying Mg concentrations, the temperature of the Mg effusion cell is adjusted between 500°C and 530°C. based on Table I, the growth is carried out using an RF power of 400W for oxygen plasma. A constant oxygen flow rate of 3ml/min, and the substrate temperature of 360°C[32].

2.2. Experimental

2.2.1. XRD and AFM

The CdMgO ternary alloy films were investigated utilizing a PANalytical X'Pert Pro MRD X-ray diffraction (XRD) equipment employing Cu K$\alpha$1 radiation (1.5406 Å wavelength). An analyzer with a 2-bounce Ge (200) hybrid monochromator is part of the diffractometer. We compared the CdMgO samples series using the XRD method by looking at the θ-2θ scans of each sample. The Bragg angle, represented by θ, the angle formed by the sample plane and the XRD source. The angle between the projection of the XRD source and the detector is 2θ. As a result, the XRD patterns generated by the geometry are frequently referred to as 2θ[32]. The surface morphology of the alloys was examined using atomic force microscopy (AFM) in tapping mode with a Bruker Dimension Icon model.

Table 1: Sample details, including effusion Cd and Mg cell temperature (T) and thickness

| Sample Ref. | T (Cd) (°C) | T (Mg) (°C) | Thickness (nm) |
|---|---|---|---|
| S1 | 380 | - | 300 |
| S2 | 380 | 500 | 300 |
| S3 | 380 | 510 | 300 |
| S4 | 380 | 520 | 300 |
| S5 | 380 | 530 | 300 |

2.2.2 Photothermal Infrared Radiometry

The experimental setup shown in Figure. 1 use a DPSS laser with a wavelength of 532nm and DC power of 800 mW[18]. For laser intensity modulation, an acoustic-optical modulator (AOM) that operates in the 100Hz to 2 MHz frequency range is employed. Two parabolic mirrors (off-axial, Au coated) are used to capture and collimate the IR radiation that the sample emits. The IR radiation is directed towards the mercury cadmium telluride (MCT) detector through a window that is coated with Germanium antireflection material. The signal is examined by lock-in-amplifier (LIA)[18].

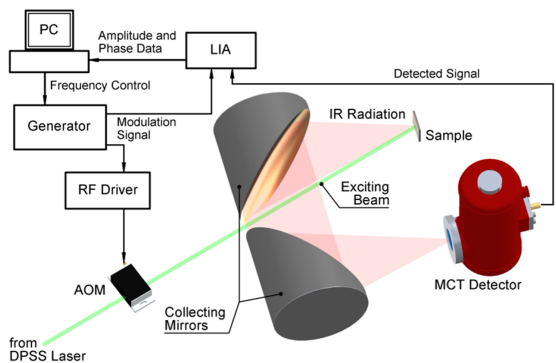

Figure 1. The IR experimental set-up used in the study [18].

2.2.3 Thermoreflectance

The experimental setup of Frequency-Domain Thermoreflectance (FDTR) shown in figure 2 enables precise measurement of thermal transport properties by analyzing modulated heating and reflectance changes. A 532 nm pump laser is electronically modulated using a generator signal between 100 Hz and 1.5 MHz frequencies, while a 488 nm probe laser detects the resulting reflectance changes. Beam splitters and a dichroic mirror are employed to align both lasers along the same optical axis. A high precision objective lens is then used to focus the lasers onto the sample, guaranteeing a Gaussian intensity distribution and spot diameters ranging from 2 μm to 4 mm. After being directed towards a photodiode, the reflected beams are transformed for phase and amplitude components[6].

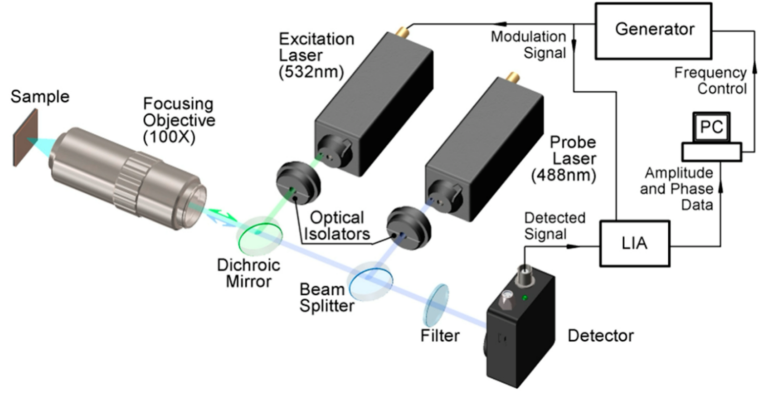

Figure 2: Schematic for the FDTR setup [6].

2.3. Theoretical model for PTR and FDTR results

The theoretical framework used in this work has already been published. Both individual layers and interfaces are taken into account in the established models and literature on multi-layer thermal transport parameters, which can be used to represent the 3D signal. In particular, these previously described models can be used to express the 3D signal in terms of Hankel variable ($\lambda$)[6]. The following notation for the multilayer coefficients can be used to express the thermoreflectance (TR) signal:

$$\theta_{TR} = \frac{-Q}{2\pi} \int_0^\infty \frac{D}{C} e^{-\frac{(\lambda d)^2}{8}} \lambda \, d\lambda \qquad (1)$$

where $\lambda$ represents Hankel variable, Q is the input pump beam intensity, $d$ signifies the beam spot, and $C$ and $D$ are coefficients known as:

$$\begin{bmatrix} A_n & B_n \\ C_n & D_n \end{bmatrix} = M_3 \, M_{32} \, M_2 \, M_{21} \, M_1 \qquad (2)$$

Where $M$ is defined as an interface to layers, $M_3$ for the third layer $M_{32}$ for between third and second layer, and so on.

$$A_n = \cosh(\sigma_n l_n) = D_n \,;\, B_n = \frac{\sinh(\sigma_n l_n)}{k_{\perp n} \sigma_n}; C_n = -k_{\perp n} \sigma_n \sinh(\sigma_n l_n) \qquad (3)$$

Where $k_\perp$ represents cross-plane thermal conductivity, $\sigma$ denoted thermal diffusivity and $l_n$ signifies the layer thickness of the nth layer, with the $\sigma^2 = \frac{k_\parallel}{k_\perp}(\lambda^2 + i\omega\rho C)$ where $k_\parallel$ denotes in-plane thermal conductivity, $\omega$ represents the pump modulation frequency, $\rho C$ signifies the volumetric specific heat.

The thermal boundary resistance at the interfce, R determined by:

$$M_{n,n+1} = \begin{bmatrix} 1 & -R_{n,n+1} \\ 0 & 1 \end{bmatrix} \qquad (4)$$

Where $R_{n,n-1}$ denotes the thermal interface resistance between the nth and the (n-1)th layers.

For the case of PTR,

$$\theta_{PTR} = -\frac{D}{C} , \qquad (5)$$

C and D is defined by formula (3) but with different value of $\sigma^2 = \frac{1}{k_\perp}(i\omega\rho C)$ :

3. Experimental results and discussion

3.1 Film crystallinity

The structural characteristics of CdMgO alloys formed on quartz surface were examined using XRD analysis. The XRD patterns were recorded over a 2θ range from 20 to 100 degrees. The diffractograms confirmed the formation of cubic rock-salt (RS) structured films without the presence of any secondary phases or impurities. For the samples consisting purely of CdO, the characteristic diffraction peaks from reflections of the (111) and (200) atomic planes are observed at diffraction angles around 33° and 39°, respectively. These peaks indicate the polycrystalline nature of the cubic RS structure of CdO, aligning well with the standard XRD data (JCPDF card no. 652908). Notably, the XRD analysis revealed no peaks associated with the quartz substrate.

The CdMgO alloys exhibit a regular shift towards greater diffraction angles for the 111 and 200 diffraction peaks as the Mg concentration rises. Previously it has been reported that films grown at lower temperatures tend to orient along the (111) direction, whereas those deposited at higher temperatures favor the [200] orientation, with this behavior typically observed around 400 °C [32]. This investigation also shows that the alloy Mg content affects the crystallographic orientation. The diffraction peaks broaden as the Mg incorporation rises, which may be related to an unequal distribution of Mg over the depth of the layers or the presence of nonhomogeneous strain distribution. Figure 3 shows the (111) and (200) diffraction peaks were fitted using a Voigt profile, from which the FWHM of the peaks (β) was extracted to calculate the crystalline domain (grain) using the Scherrer equation (grain size, $\tau = k\lambda/(\beta \cos\theta)$; k = Scherrer constant (0.94), $\lambda$ = X-ray wavelength used, and $\theta$ = diffraction angle). The obtained grain size is listed in Table 2.

Table 2: CdMgO crystalline grain size obtained from the fit of the (111) and (200) diffraction peaks

| Sample Ref. | 111 (nm) | 200 (nm) |
|---|---|---|
| S1 | 41 | 72 |
| S2 | 16 | 17 |
| S3 | 19 | 11 |
| S4 | 11 | 9 |
| S5 | 4 | 4 |

Previous research by Adhikari et al.[33] highlighted the growth of double XRD peaks in $Cd_xMg_{1-x}O$ films grown on m- and c-plane sapphire substrates using the plasma-assisted MBE technique. This result showed that the formation of regions with different concentrations within the CdMgO ternary layers, showing that both the growth conditions and the substrate orientation greatly affect the uniformity of the alloy layers.

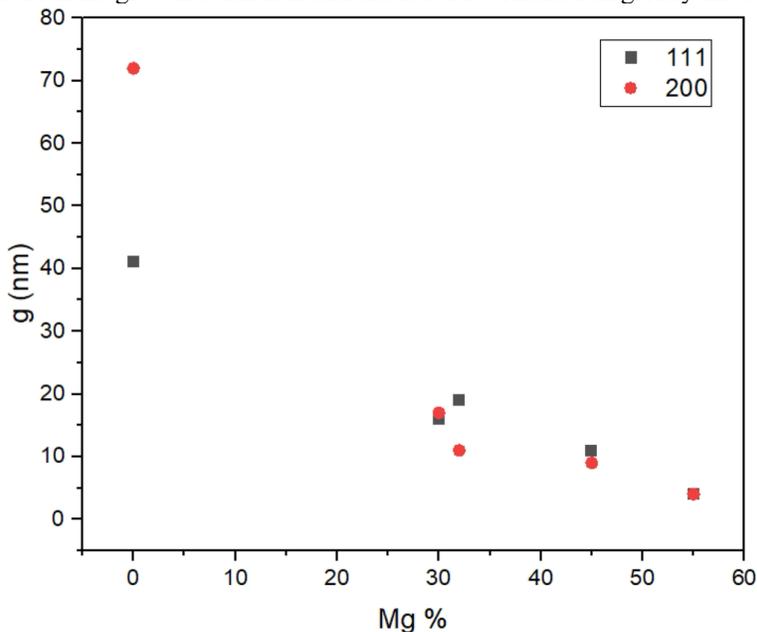

Figure 3: Crystallite grain size (g) and **FWHM** of the main (111) and (200) diffraction peaks.

3.2 Surface Morphology

The AFM results, as depicted in figure 4, provide a detailed analysis of the surface morphology of CdMgO ternary alloys with varying magnesium (Mg) content. The AFM images labeled S1 through S5 show the surface topography of the samples over a 5x5 μm$^2$ area, with a vertical scale of 50 nm. These images reveal differences in the surface roughness and texture among the samples, which are quantitatively expressed as average roughness ($R_a$) values in table 3. The roughness values $R_a$ correlate with the Mg content in the samples. The sample with the highest Mg concentration has the smallest $R_a$ value. The smoothing of the surface morphology due to the change in the alloys Mg concentration is implied by the drop in the average $R_a$. The average roughness parameter value for the whole Mg concentration range of CdMgO alloys varies between 2 to 10 nm.

These AFM results demonstrate that the surface roughness of CdMgO ternary alloys can be finely tuned by adjusting the Mg content, which is crucial for optimizing the light scattering and trapping properties in photovoltaic applications[34]. The trend observed here is consistent with previous studies, where smoother surfaces were reported for higher Mg concentrations in CdO:Mg films prepared by different methods on various substrates. The ability to control the surface morphology and roughness through compositional changes is an important aspect of material design for enhancing the efficiency of solar cells.

Table 3: Composition and roughness values of the CdMgO films.

| Sample Ref. | Mg Content (mole fraction) | $R_a$ (nm) |
|---|---|---|
| S1 | 0 | 13.52 |
| S2 | 27 | 2.10 |
| S3 | 30 | 8.72 |
| S4 | 45 | 1.14 |
| S5 | 55 | 0.93 |

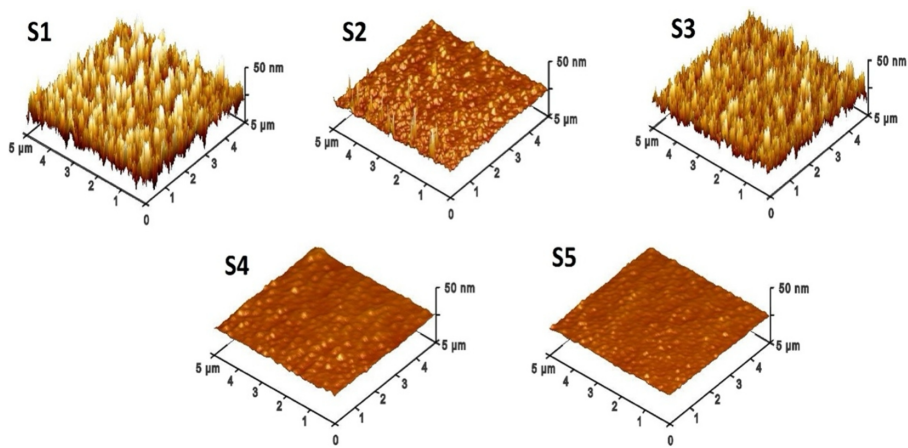

Figure 4: Average roughness of the CdMgO films deposited on quartz substrates as a function of Mg content.

3. 3. Photothermal infrared radiometry

Figure 5 shows the normalized PTR amplitude and phase for CdMgO (sample S2). Other samples shown the same findings. PTR amplitude and phase was normalized using 5 mm thick glass carbon.

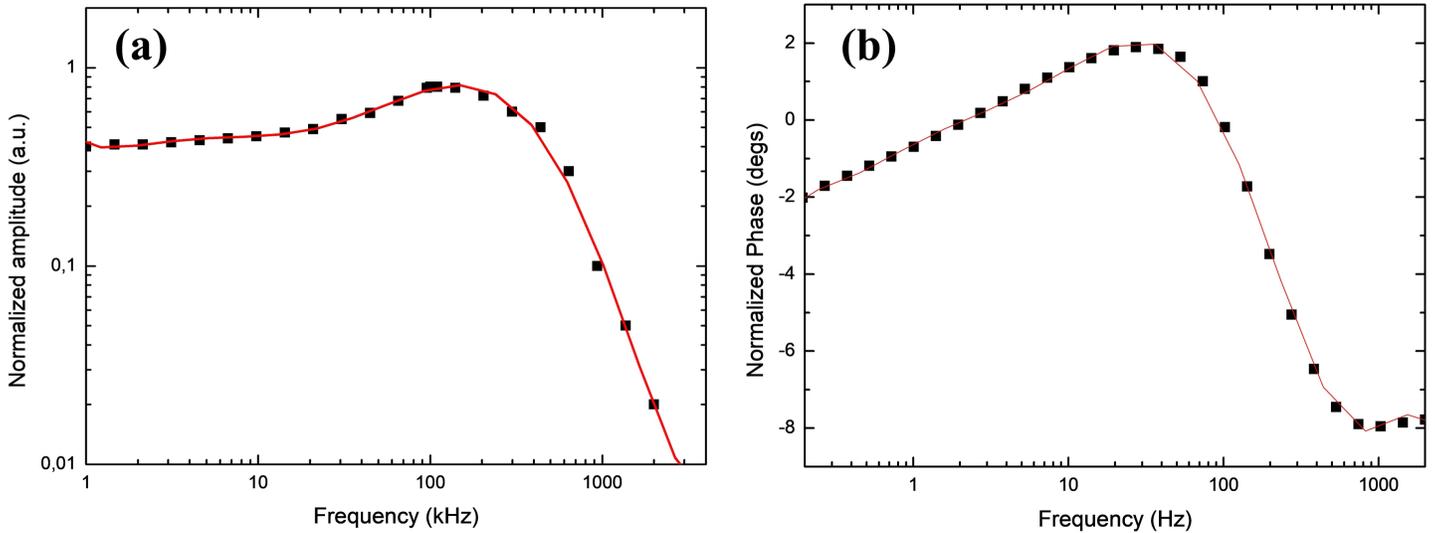

Figure 5: Variation of the PTR (a) amplitude ratios and 5(b) phase difference with frequency for the case of the CdMgO 27 mole fraction of Mg (S2).

3.4. Thermoreflectance

Figure 6 shows the normalized TR amplitude and phase for CdMgO (sample Z17). Other samples shown the same findings. TR amplitude and phase was normalized using photodiode placed instead of sample.

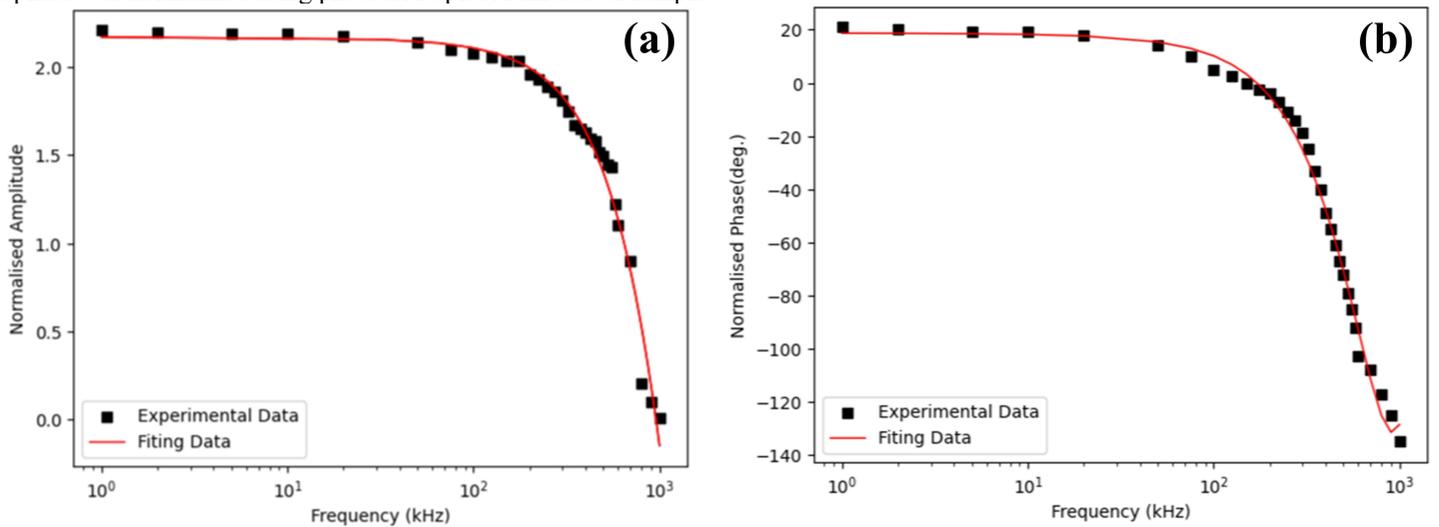

Figure 6: Variation of the FDTR (a) amplitude ratios and 6(b) phase difference with frequency for the case of the CdMgO 30 mole fraction of Mg (S3).

s

Table 4: Estimated parameters using PTR and TR methods.

| Mg. Conc. (mole fraction) | $k_\perp$ (W/mK) | $\sigma_\perp \times 10^{-6}$ (m²/s) | $k_\parallel$ (W/mK) | $R_{th} \times 10^{-8}$ (m²W/K) | η |
|---|---|---|---|---|---|
| 0 | 7.98±0.13 (PTR)<br>7.67±0.14 (FDTR) | 3.50±0.07 | 15.35±0.32 (FDTR) | 6.06±0.3 (PTR)<br>6.07±0.2 (TR) | 2.00214 |
| 0.27 | 3.95±0.08 (PTR)<br>3.87±0.09 (FDTR) | 2.20±0.07 | 4.21±0.09 (FDTR) | 4.22±0.12 (PTR)<br>4.01±0.15 (FDTR) | 1.08810 |
| 0.3 | 4.20±0.08 (PTR)<br>4.21±0.08 (FDTR) | 2.60±0.06 | 2.33±0.03 (FDTR) | 3.17±0.07 (PTR)<br>3.44±0.03 (FDTR) | 0.55315 |
| 0.45 | 2.75±0.08 (PTR)<br>2.88±0.07 (FDTR) | 2.80±0.08 | 1.71±0.04 (FDTR) | 4.68±0.14 (PTR)<br>4.46±0.17 (FDTR) | 0.58844 |
| 0.55 | 2.87±0.09 (PTR)<br>2.76±0.08 (FDTR) | 4.70±0.07 | 0.56±0.08 (FDTR) | 5.20±0.3 (PTR)<br>4.99±0.4 (FDTR) | 0.20200 |

## 4. Discussion

The thermal conductivity of CdMgO thin films as a function of magnesium content ($x$), with the best fit with the Abeles model (eq 7) [31]:

$$k_{CdMgO} = \left[\frac{1-x}{k_{CdO}} + \frac{x}{k_{MgO}} + cx(1-x)\right]^{-1} \qquad (7)$$

where $k$ is the cross-plane thermal conductivity, $c$ represents the disorder in the alloy system, $x$ is Mg %, $k_{CdO}$ and $k_{MgO}$ are thermal conductivity of Cd and Mg respectively. Figure 7 shows the best fit to the Abeles model. The model captures the effects of phonon-interface scattering and thermal boundary resistance, highlighting the impact of material composition on phonon transport [35].

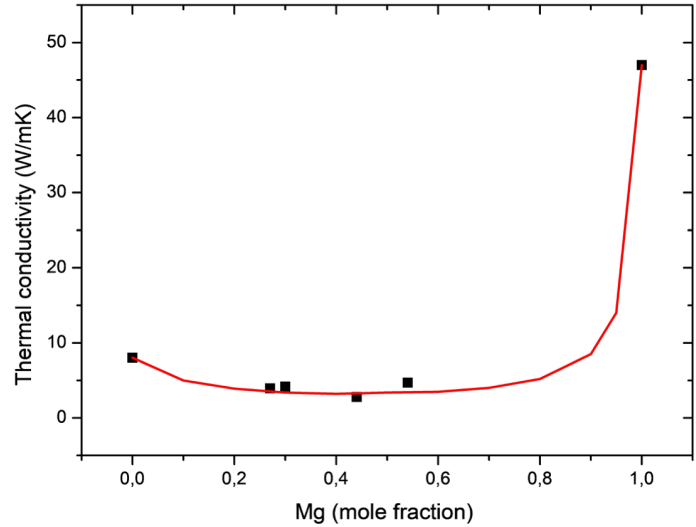

Figure 7: Cross-plane thermal conductivity as a function of magnesium content with best fit to Abeles model.

Figure 8 demonstrates the relationship between the thermal conductivity and the crystalline grain size in CdMgO alloy films. The graphic highlights the effectiveness of the Callaway model, which is expressed in equation (8) [30] in forecasting the thermal behaviour by comparing experimental data with the model fit. The data shows that the cross-plane thermal conductivity tends to increase linearly with grain size.

$$k = CT^3 \int_0^{\theta_D/T} \frac{x^4 e^x (e^x-1)^{-2}}{\alpha T^4 x^4 (\beta_1+\beta_2) T^5 x^2 + v/L} dx + k_2 \quad (8)$$

Where C can be expressed as:

$$C = \frac{k_B}{2\pi^2 v}\left(\frac{k_B}{h}\right) \quad (9)$$

And x can be expressed as $x = \frac{\hbar \omega}{k_B T}$

$T$ represents the absolute temperature, $\theta_D$ represents the Debye temperature, $v$ represents the phonon velocity, $L$ represents a characteristic length of the material, $k$ denoted as thermal conductivity, $\omega$ represents the circular frequency, $k_B$ represents the Boltzmann constant, and $h$ represents the Planck constant. $\beta_1+\beta_2 T^5 x^2$ for boundary and impurity scattering.

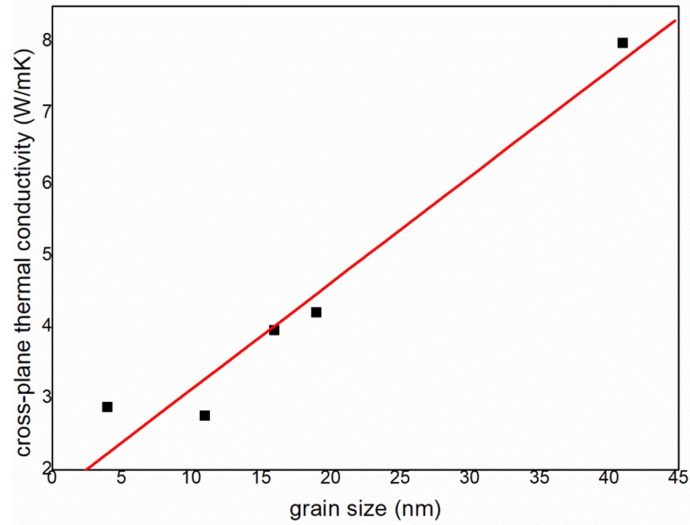

Figure 8: Cross-plane thermal conductivity vs crystalline grain size.

Finally, the volumetric heat capacity is plotted as a function of the magnesium concentration in Figure 9. The volumetric heat capacity decreases with the increase in Mg concentration. This can be explained in the frame of Debye model, where the phonon contribution to heat capacity is:

$$C_V = 9Nk_B \left[\frac{T}{\theta_D}\right]^3 \int_0^{x_D} \omega \frac{x^4 e^x}{(e^x-1)^2} dx \quad (10)$$

Where $x = \frac{\hbar \omega}{k_B T}$ \quad (11)

Where, $C_V$ Heat capacity at constant volume, $N$ is number of atoms, $k_B$ is denoted as Boltzmann constant, $T$ is room temperature $\theta_D$ is Debye temperature of material and $x$ is phonon frequency normalized by temperature.

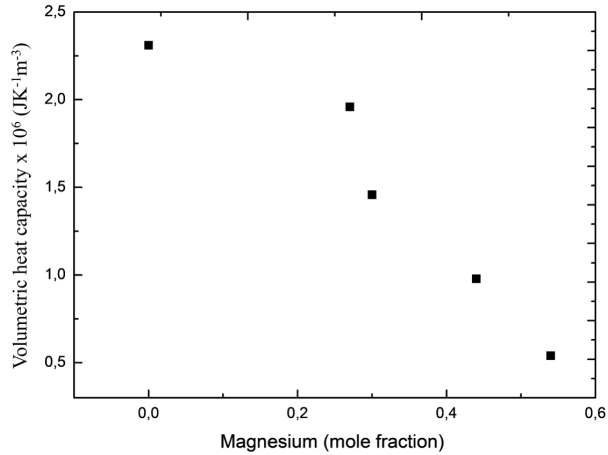

Figure 9: Measured Volumetric heat capacity as a function of magnesium concentration, respectively.

As one can see from Figure 9, measured and predicted values for the heat capacity of CdMgO should decrease with increasing magnesium concentration. According to Debeye law, if we have two different materials such as CdO and MgO, their specific heat is higher for CdO than for MgO because the Debeye temperature of CdO 255 K [36] and while for MgO is 755 K [37].

### 4.1. Effect of Surface Roughness on Thermal Conductivity of CdMgO

The thermal conductivity of CdMgO alloys is strongly influenced by surface roughness, as shown in Figure 10, which affects phonon scattering mechanisms critical for heat transport. AFM analysis (Figure 4) shows that increasing magnesium content reduces surface roughness, from 13.5 nm for pure CdO to 0.9 nm at 55 at.% Mg, as summarized in Table 3. Previously Martin et al. [38] plotted thermal conductivity as a function of nanowire diameter for various roughness $R_a$ values. They showed that as the diameter decreases, the thermal conductivity significantly declines, particularly for nanowires with higher surface roughness. This trend underscores the substantial impact of surface roughness on thermal transport, especially in smaller diameter nanowires. Higher surface roughness amplifies phonon scattering, particularly at grain boundaries, leading to an increase in in-plane thermal conductivity, as shown by the black points in Figure 10. The dependence of in-plane and cross-plane thermal conductivities on $D/R_a^2$ demonstrates that in-plane thermal conductivity decreases as $D/R_a^2$ increases, with a strong fit to defect scattering models. Cross-plane thermal conductivity, shown by the red points in Figure 10, follows a similar trend but remains consistently lower due to anisotropic transport, attributed to structural factors like layered morphology and weaker interlayer bonding.

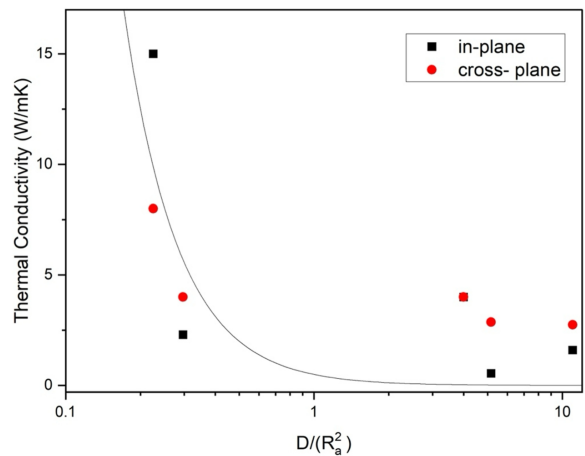

Figure 10: Effect of ratio of average grain size and square of the average surface roughness on the in-plane (black dots) and cross-plane (red dots) thermal conductivity of CdMgO films.

## 4.2. Inverse Thermal Anisotropy in CdMgO Thin Films

The thermal conductivity measurements of disordered CdMgO thin films as a function of varying magnesium concentration is shown in Figure 10. The graph likely presents both in-plane and cross-plane thermal conductivity values, highlighting the phenomenon of inverse thermal anisotropy, where the cross-plane thermal conductivity is greater than the in-plane conductivity. This unusual behavior is significant for applications in optoelectronics and thermoelectrics, as it suggests that heat can be dissipated more effectively in the direction perpendicular to the film surface compared to along the surface itself. The data may reveal how increasing Mg concentration influences thermal transport properties, potentially correlating with structural changes observed in previous analyses, such as grain size and surface morphology. The findings highlight the necessity of changing Mg content to improve thermal management features, essential for augmenting device performance. Figure 11 serves as a crucial visual representation of the findings regarding the distinct thermal properties of CdMgO thin films. Figure 12 illustrates the inaccuracies in the estimation of thermal anisotropy.

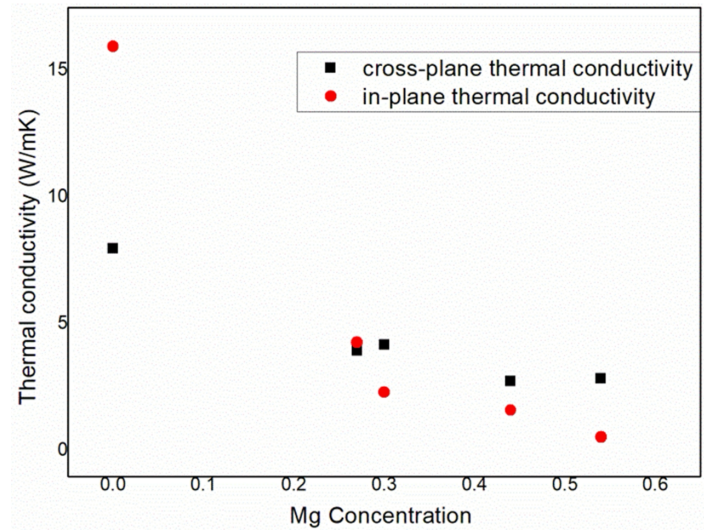

Figure 11: Thermal Anisotropy behaviour of in-plane and cross-plane thermal conductivity of CdMgO thin films as a function of Mg concentration.

Errors of estmination thermal anisotropy are shown on Figure 12

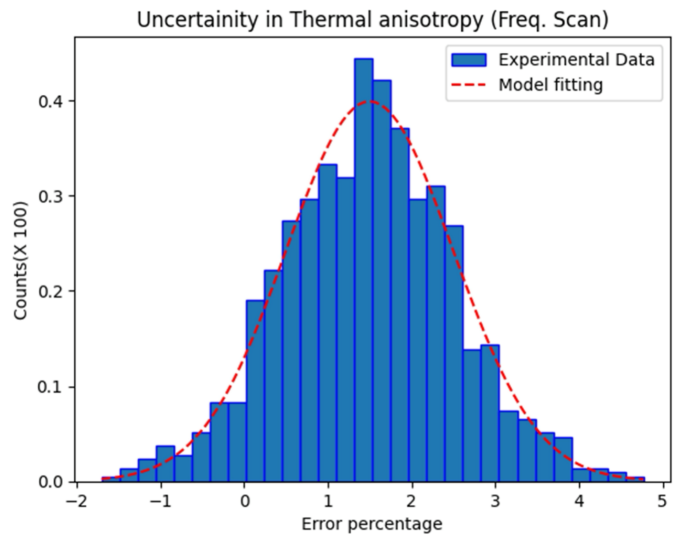

Figure 12: Errors of estimation thermal anisotropy in figure 11

## 5. Conclusion

This study revealed the inverse thermal anisotropy exhibited by CdMgO thin films, as characterized by photothermal infrared radiometry and thermoreflectance analysis, demonstrating their superior thermal transport characteristics. With increasing Mg concentration, XRD analysis revealed a systematic shift of diffraction peaks toward higher *2θ* angles accompanied by peak broadening, indicating lattice contraction and reduced crystallinity. This structural evolution was further corroborated by AFM analysis, which showed that increasing Mg content significantly improved surface smoothness, with the average surface roughness decreasing from 13.52 nm for pure CdO to 0.93 nm at 55% Mg concentration.

Using PTR and FDTR, important thermophysical properties were measured. The cross-plane thermal conductivity ($k_\perp$) and in-plane thermal conductivity ($k_\parallel$) values across different Mg concentrations were obtained. For instance, in the sample with 0% Mg, $k_\perp$ was measured as $7.98\pm0.13$ Wm$^{-1}$K$^{-1}$ by PTR and $7.67\pm0.14$ Wm$^{-1}$K$^{-1}$ by FDTR, while $k_\parallel$ was $15.35\pm0.32$ Wm$^{-1}$K$^{-1}$ (FDTR). These results clearly showed the inverse thermal anisotropy in CdMgO thin films, where $k_\perp > k_\parallel$. The thermal diffusivity ($\sigma_\perp$) also varied with Mg concentration, such as $3.50\pm0.07\times10^{-6}$ m²s$^{-1}$ in the 0% Mg sample.

In addition to the experimental findings, the thermal behavior of the CdMgO thin films was further interpreted through theoretical modeling using the Abeles and Callaway models. The Abeles model effectively described the impact of compositional variation on phonon scattering and thermal transport, while the Callaway model successfully captured the observed linear relationship between cross-plane thermal conductivity and crystalline grain size, providing strong theoretical support for our experimentally achieved results.

Furthermore, according to the Debye model, the volumetric heat capacity decreased with increasing Mg concentration, which can be attributed to the difference in Debye temperatures between CdO (255K) and MgO (755K). Surface roughness was found to significantly influence thermal conductivity, with higher roughness enhancing phonon scattering at grain boundaries. This scattering effect impacted both in-plane and cross-plane thermal conductivities. Specifically, the in-plane thermal conductivity decreased as the ratio of average grain size to the square of the average surface roughness ($D/R_a^2$) increased, following a trend consistent with defect-scattering models.

In summary, this study comprehensively examines the thermal properties of CdMgO thin films, providing valuable insights for optimizing CdMgO-based devices. Future research should explore the correlation between microstructure, thermal properties, and device performance, as well as the potential of thin films in applications requiring precise thermal control.

## Acknowledgements

The work was supported by the Polish NCN Projects DEC-2021/41/B/ST5/00216.